\documentclass[usenatbib,usegraphicx]{mn2e}

\title[The elliptical galaxy CMR as a discrminant between galaxy formation paradigms]
	{The elliptical galaxy colour-magnitude relation as a discriminant between the
	monolithic and merger paradigms}

\author[S.~Kaviraj et al.]
{S. Kaviraj$^{1}$\thanks{E-mail: skaviraj@astro.ox.ac.uk},
  J. E. G. Devriendt$^{1,2}$, I. Ferreras$^{1,3}$ and S. K. Yi$^{1}$\\
$^{1}$Department of Physics, University of Oxford, Keble Road,
  Oxford OX1 3RH, UK\\
$^{2}$Observatoire Astronomique de Lyon, 9 Avenue Charles Andr\'e,
  69561 Saint-Genis Laval cedex, France\\
$^{3}$Department of Physics, Institute of Astronomy, ETH Hoenggerberg
  HPF D8, 8093 Zurich, Switzerland}

\begin{document}

\date{14 July 2003}

\pagerange{\pageref{firstpage}--\pageref{lastpage}} \pubyear{2003}

\maketitle

\label{firstpage}

\begin{abstract}
The colour-magnitude relation (CMR) of cluster elliptical galaxies
has been widely used to constrain their star formation histories (SFHs) and to
discriminate between the monolithic collapse and merger paradigms of elliptical
galaxy formation. We use a $\Lambda$CDM
hierarchical merger model of galaxy formation to investigate the
existence and redshift evolution of the elliptical galaxy CMR in the
merger paradigm. We show that the star formation
history (SFH) of cluster ellipticals predicted by the model is
\emph{quasi-monolithic}, with only $\sim$ 10
percent of the total stellar mass forming after z $\sim$ 1. 
The quasi-monolithic SFH results in a predicted CMR that agrees well
with its observed counterpart in the redshift range $0<z<1.27$. We use
our analysis to argue that the \emph{elliptical-only} CMR can be used
to constrain the SFHs of present-day cluster ellipticals 
\emph{only if we believe a priori in the monolithic collapse
model}. It is not a meaningful tool for constraining the SFH in
the merger paradigm, since a progressively larger fraction of the
progenitor set of present-day cluster ellipticals is contained in
late-type star forming systems at higher redshift, which cannot be
ignored when deriving the SFHs. Hence, the elliptical-only CMR is not a useful
discriminant between the two competing theories of elliptical galaxy
evolution.
\end{abstract}


\section{Introduction}
Evolution with redshift of fundamental physical relations can
provide robust constraints on the epoch of formation and
subsequent evolution of early-type galaxies. The apparently
universal relationship between colour and luminosity  of
elliptical galaxies, usually referred to as the \emph{colour-magnitude
  relation} (CMR), was first established by \citet{SV77},
although the correlation between these two quantities had been
demonstrated before
\citep[e.g.][]{Baum59,deVaucouleurs61,McClure68}. The observed CMR has
been widely used as a discriminant between the two competing
theories of early-type galaxy evolution, the monolithic collapse
model \citep[e.g.][]{Larson74,KA1997} and the
hierarchical merger model \citep[e.g.][]{KWG93,SP99,Cole2000,Hatton2003,Khochfar2003}.

A comprehensive study of the CMR, using photometric data based on
CCD observations of the nearby Virgo and Coma clusters, was first
undertaken by \citet*[hereafter BLE92]{BLE92}. Their results showed
a remarkably small scatter about the mean relation. Their
interpretation of the results, \emph{in the context of the
monolithic collapse model}, was to attribute the slope of the
CMR to a variation in mean metallicity with luminosity and to
attribute the small scatter to a small age dispersion between
galaxies of the same size. They concluded that the epoch of
formation of elliptical galaxies should be at $z>2$. Subsequent
studies of the CMR extended the BLE92 results to intermediate
redshifts ($0<z<1$) and showed that there was no detectable
evolution of the slope and scatter with time
\citep[e.g.][]{Ellis97,Stanford98,Gladders98,VD2000}. The results
from these studies were interpreted as confirmation of a high-redshift
formation epoch of cluster ellipticals followed by
passive evolution to present day.

Subsequent studies indicated that the key characteristics
of the CMR (slope and scatter) that were derived by these
authors needed some modification. Elliptical galaxies commonly display
radial colour gradients
\citep[e.g.][]{deVaucouleurs61,SV78,FIH89,Peletier90}, being optically
redder at their cores than at the outskirts. Galaxy
colours in the majority of these CMR studies were derived using
\emph{fixed apertures} which, given that a galaxy's intrinsic size
may vary, meant sampling different portions of different galaxies. Accounting for the
effect of fixed aperture photometry is essential in the context of this
study, since our model lacks spatial information on the scale of
galaxies. Producing a \emph{central} CMR is therefore not possible
within the model and we must compare our results to aperture-corrected
photometry.

Numerous authors have attempted to quantify the \emph{fixed-aperture
bias} that may result due to the presence of colour
gradients. An efficient way to control this effect is to measure the
colour inside an aperture which scales with the size of the galaxy
\citep[e.g.][]{Bower98,Terlevich2001,Sco2001}, such that one samples an
identical fraction of the light in each galaxy. \citet{Bower98}
compared the CMR slope observed by BLE92 with the slope derived after
replacing a fixed aperture with the parameter $D_v$ - the size of the
galaxy within which the mean surface brightness is
19.80 mag arcsec$^{-2}$. They estimated that colour gradients
accounted for roughly 30 percent of the slope i.e. the magnitude of
the slope in the $D_v$ CMR was roughly two-thirds of that derived using fixed-apertures
in BLE92. Crucially, their study indicated that the CMR maintains a
\emph{significant slope} even after correcting for colour gradients. A series of
other studies have also studied the effect of removing the
fixed-aperture bias. Shallower CMR slopes have been reported
by \citet{Prug96} who used colours derived within the \emph{effective
  radius} (radius which contains half the galaxy's light, $R_e$)   
and by \citet{Fioc99} who used total magnitudes
and colours. \cite{Sco2001}
suggested that re-computing colours over $R_e$ for the BLE92 galaxies
causes the apparent slope to decrease from
$-0.082 \pm 0.008$ to $-0.016 \pm 0.018$, a value which is
statistically consistent with a zero slope. In addition, the scatter
increases from $0.035$ to $0.136$, due to the large intrinsic
scatter in the colour gradients. However, the uncertainty in colour
measurements within $R_e$ can be significantly larger than within $D_v$,
because the surface brightness within $D_v$ is higher. At $2\sigma$
the \citet{Sco2001} data is consistent with slopes between $+0.02$ and
$-0.52$. Modifying the BLE92 slope for colour gradients, using the 30
percent correction factor derived by \citet{Bower98}, gives $-0.054$,
so that within the errors there is agreement between the various studies. 

Over the last decade, there has been steadily accumulating evidence
for morphological evolution amongst
cluster galaxies, which suggests that formation mechanisms of cluster
ellipticals are at
least \emph{not uniquely monolithic}. Although approximately 80
percent of galaxies in the cores of present day clusters have
early-type morphologies \citep{Dressler80}, a higher fraction of
spiral galaxies have been reported in clusters at $0.3<z<0.8$
\citep[e.g.][]{BO84,Dressler97,Couch98,VD2000}, along with increased
rates of merger and interaction events
\citep[e.g.][]{Couch98,VD99}. \citet{KCW96} suggested that only
approximately one-third of early-type galaxies in the
Canada-France Redshift Survey \citep{Schade1995} were fully formed and evolving
passively. \citet{Franceschini98} found a remarkable absence of
early-types galaxies at $z>1.3$ in a K-band selected sample in the
Hubble Deep Field. These
results strongly suggest that early-type galaxies in nearby and
distant clusters may have been formed from late-type progenitors
\citep[e.g.][]{BO84,Dressler97} and highlight the possible if not
essential role of merger and interaction events in the formation
of early-type galaxies. In particular, if the merger paradigm is
correct then late-type progenitors of the present-day cluster
ellipticals must be included in any method (e.g. the CMR) employed
to determine their SFHs. Excluding these
late-type progenitors would produce a distorted view of their
formation histories (progenitor bias), a point first suggested
and explored in detail by \citet{VD2001a}. 

Given the accumulating evidence for formation of early-type galaxies
from star-forming progenitors at fairly recent epochs, a number of
authors have successfully reconciled the observed CMR with galaxy
merging models \citep{KC98,Bower98,Shioya98,VD98}. Apart from
\citet{KC98}, these studies have not involved a fully realistic
semi-analytical galaxy formation model which incorporates the
important effects of galaxy merging on the chemo-photometric evolution
of galaxies. One of our aims is to extend these studies by applying a $\Lambda$CDM
hierarchical merger model to study the CMR from low to high
redshift.

We begin our study by discussing the comparative effects of age
and metallicity in determining the model $(U-V)$ CMR at
present day and tracing the bulk SFHs of
cluster ellipticals as a function of redshift. We then explore
the predicted evolution of the CMR to high redshifts, compare with
existing observational evidence and, in particular, quantify the
effect of progenitor bias. Using our analysis
of progenitor bias, we present arguments to show that the commonly
used \emph{elliptical-only} CMR, even when it is derived using
equal light fractions \citep[c.f.][]{Bower98,Terlevich2001,Sco2001},
can only be used to constrain the SFHs of cluster
ellipticals \emph{if we believe a priori in a monolithic collapse
model}. It is not a meaningful method of constraining the SFH in
the hierarchical merger picture. Hence it is also not a useful
discriminant between the two competing theories of galaxy
evolution.

\section{Model parameters that affect the present-day CMR}
The model we use in this study is GALICS, which combines
large scale N-body simulations with simple analytical recipes for
the dynamical evolution of baryons within dark matter haloes. We
direct readers to \cite{Hatton2003} for specifics regarding the
model. There are certain key parameters in the model that affect the age
and metallicity of the model galaxies and thus have an impact on
the slope, scatter and absolute colour of the predicted CMR. A
discussion of these model parameters is necessary, not only to
elucidate their effect on the CMR, but also because the actual
setup we use in this study is slightly different from the
reference model given in \citet{Hatton2003}. The setup has been
altered, firstly to make some corrections to the metallicity of
fresh gas injected into DM haloes, and secondly to bring the
predicted metallicities of the model galaxies in agreement with
current observational evidence. Table \ref{tab:model_params}
summarises the changes in the characteristics of the CMR due to
variations in these parameters. In the subsequent sections we
present an explanation of the parameters and the values used in
this study.

\begin{table*}
\begin{center}
\begin{minipage}{126mm}
\caption[Baseline metallicity, threshold black hole mass and IMF]
{\small{Variations in CMR slope and scatter with baseline
    metallicity, threshold BH mass and IMF. The slope is derived from
    least squares fits (over the magnitude range $M_V = -19$ to $M_V =
    -23$) and the scatter is calculated using Tukey's
    bi-weight statistic. The values in bold indicate the parameters used 
    for this study. $^{\dagger}$BHT = Black Hole Threshold mass.}}
\label{tab:model_params}
\begin{tabular}[width=3.4in]{c|c|c|c}

    \multicolumn{4}{c}{Baseline metallicity = 0}\\
    & BHT = 45$M_{\odot}$ & BHT = 60$M_{\odot}$ & BHT = 120$M_{\odot}$\\
    \hline \hline

           &   &   & \\
Kennicutt IMF  & $-0.037 \pm 0.007$ & $-0.049 \pm 0.007$ & $-0.050 \pm 0.009$ \\
           & $0.072$            & $0.079$            & $0.094$            \\ \hline
           &   &   & \\
Scalo IMF      & $-0.034 \pm 0.006$ & $-0.033 \pm 0.009$ & $-0.047 \pm 0.010$ \\
           & $0.057$            & $0.070$            & $0.10$             \\

    \hline \hline
    \multicolumn{4}{c}{}\\
    \multicolumn{4}{c}{Baseline metallicity = 0.1$Z_{\odot}$ }\\
    & $\textnormal{BHT}^{\dagger}$ = 45$M_{\odot}$ & BHT = 60$M_{\odot}$ & BHT = 120$M_{\odot}$\\
    \hline \hline

           &   &   & \\
Kennicutt IMF  & $-0.036 \pm 0.009$ & $\mathbf{-0.047 \pm 0.010}$ & $-0.052 \pm 0.010$ \\
           & $0.075$            & $\mathbf{0.082}$           & $0.12$             \\ \hline
           &   &   & \\
Scalo IMF      & $-0.032 \pm 0.007$ & $-0.034 \pm 0.010$ & $-0.045 \pm 0.010$ \\
           & $0.061$            & $0.078$            & $0.10$             \\

    \hline
\end{tabular}
\end{minipage}
\end{center}
\end{table*}

\subsection{Baseline metallicity}
The reference model in \citet{Hatton2003} adds \emph{pristine} i.e. metal-free
gas to DM haloes when they are identified. However, the haloes are
not identified until they achieve a threshold mass of $10^{11}
M_{\odot}$. In reality, early population II stars would already have
polluted the ISM in the time that it takes for such halo
identifications to take place. Hence, the gas in the haloes should
not be pristine but \emph{slightly polluted}. Chemical enrichment
models \citep[e.g.][]{Devriendt99} suggest that this pollution
should be of the order of $0.1 Z_{\odot}$. Hence, we use this value as a
\emph{baseline metallicity} for fresh gas injected into DM haloes in the model. The
baseline metallicity has a negligible impact on the
slope and scatter of the CMR but slightly affects the absolute colour of the cluster sample as
it changes the average metallicities of the model galaxies.

\subsection{Threshold black hole mass and IMF}
Another parameter that affects the metal input into the ISM, and
therefore the average metallicity of the stellar population, is
the threshold mass at which a star becomes a black hole (BH). This
is still poorly understood, but estimates suggest
masses around $50 \pm 10 M_{\odot}$ \citep{Tsujimoto97}, based on a combination of
local stellar [O/Fe] abundances and chemical enrichment analysis. We use
a threshold black hole mass of $60 M_{\odot}$ in this study.

Since massive stars make a significant contribution to the metal
enrichment of the inter-stellar medium (ISM), the proportion of
massive stars and hence the IMF affects the mean metallicities of
the model galaxies. In this study we use a Kennicut IMF
\citep{Kennicutt83}, which was also used in the fiducial model of
\citet{Hatton2003}. Both the BH threshold and IMF increase the dynamical
metal enrichment of the ISM. This changes not only the mean
metallicity of the galaxies, and hence their absolute colour, but
also makes the slope of the CMR steeper. This is because more massive
galaxies, which have deeper potential wells, retain gas and
therefore metals more effectively, leading to higher enrichment of
the ISM and stellar populations that are born
from it. However, less massive galaxies tend to lose their gas
content anyway, so that a larger injection of metals into their
ISM does not have a big impact on the metallicity of their stellar
populations. As a result of this differential behaviour the slope
of the CMR becomes steeper.

\section{Properties of present-day cluster ellipticals}
\begin{figure}
\begin{center}
\includegraphics[width=3.4in]{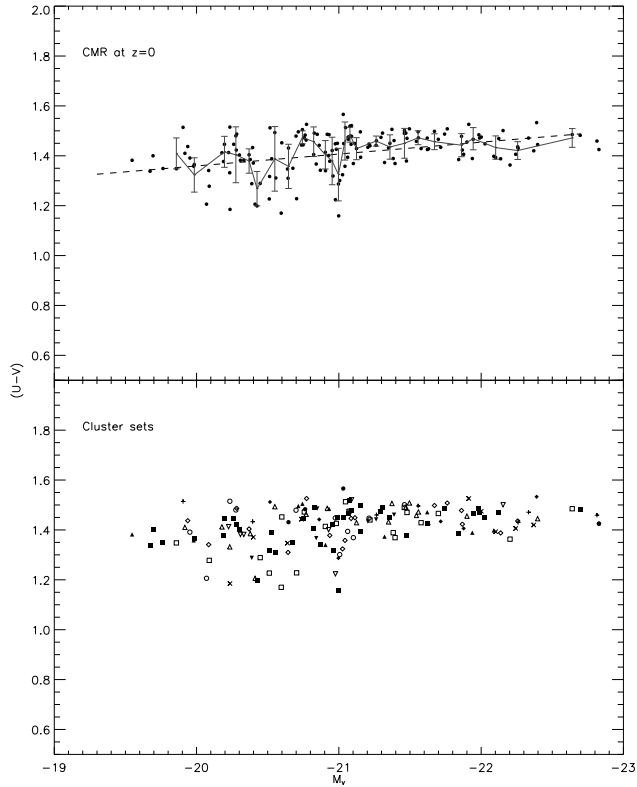}
\caption[Colour magnitude relation at $z=0$]{\small{The model colour
    magnitude relation at $z=0$. TOP: CMR sequence with a linear
    least-squares fit (dotted line) 
    and a progressive one-sigma fit to the sample. BOTTOM: Cluster
    ellipticals split into their individual clusters.}}
\label{fig:70_CMR}
\end{center}
\end{figure}

In this section, we explore the \emph{present-day} CMR predicted by
our model. Figure \ref{fig:70_CMR} presents the predicted CMR in our model
for cluster ellipticals at $z=0$. Also shown is a linear least-squares
fit to the points (dotted line) and a progressive
one-sigma fit to the sample, with the error bars indicating the
local spread of points about the best-fit relation. We select
present-day cluster ellipticals by identifying elliptical galaxies
in dark matter (DM) haloes with masses of $10^{14}M_\odot$ and
above. Also shown in Figure \ref{fig:70_CMR} is the CMR sequence
with galaxies coded according to their mean metallicities and
ages. The bottom panel splits the model
ellipticals into their individual clusters. The model slope is
derived in all cases using a linear least-squares
fit. The scatter is calculated using Tukey's Biweight statistic
\citep{Beers90}, which has commonly been used by observers in CMR
studies. Table 1 compares the model CMR with those derived by BLE92
and \citet{Bower98} within $D_v$ and by \citet{Sco2001} using the
effective radius, $R_e$.

\begin{table*}
\begin{center}
\begin{minipage}{126mm}
\caption{\small{Comparison between the characteristics of our model
    CMR at $z=0$ with BLE92, \emph{corrected for colour gradients using the 30 percent correction}
    given in \citet{Bower98} and \citet{Sco2001} who used the effective radius ($R_e$) of galaxies
    to derive colours. For the model CMR the slope is derived from
    least squares fits (over the magnitude range $M_V = -19$ to $M_V =
    -23$) and the scatter is calculated using Tukey's
    bi-weight statistic.}}
\begin{tabular}{c|c|c|l}

      Source & Slope & Scatter (mag) & Colour\\ \hline \hline
      \citet{BLE92,Bower98} (Coma)   & $-0.054 \pm 0.007$ & 0.049 & U-V (within $D_v$)\\
      \citet{Sco2001} (Coma)         & $-0.016 \pm 0.018$ & 0.136 & U-V (within $R_e$)\\
      \bf{This study}                & $\mathbf{-0.047 \pm 0.010}$ & 
      \bf{0.082} & \bf{U-V (total magnitudes)}

\end{tabular}
\end{minipage}
\end{center}
\label{tab:CMR_present_day}
\end{table*}

\begin{figure}
\begin{center}
\includegraphics[width=3.4in]{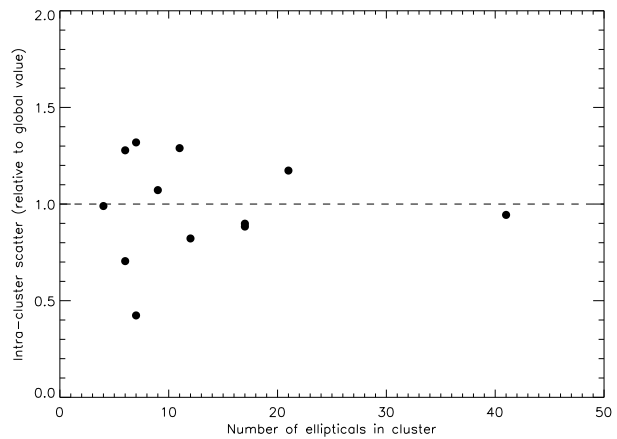}
\caption[Intra-cluster vs. total scatter]{\small{
Intra-cluster scatter as a fraction of the total scatter plotted
against the elliptical occupancy of each cluster.}}
\label{fig:cluster_scatter}
\end{center}
\end{figure}

The predicted model slope is consistent with both the value reported by
\citet{Sco2001} and the BLE92 value, after correcting for colour
gradients using the 30 percent correction derived in
\citet{Bower98}. The predicted scatter is smaller than that derived
by \citet{Sco2001} but roughly 1.5 times larger than that reported by
\citet{Bower98}. We note, however, that the scatter in the model
galaxies itself varies from cluster to
cluster, so that the \emph{intra-cluster} scatter may be different
from the global value across all clusters. In Figure
\ref{fig:cluster_scatter} we split our sample of model ellipticals
into their respective clusters and plot the intra-cluster scatter
as a fraction of the global scatter in the sample. We find, for
example, that the cluster with the largest number of ellipticals has a
lower scatter than the global value, although there are other large
clusters which exhibit a scatter above the global
value. We note that our model galaxy sample is an ensemble of galaxy sets
from different clusters, whereas the observations usually refer to only
one cluster. There may be additional issues contributing to
the discrepancy between the model and observed scatter - for example,  
there appear to be strong radial colour gradients in cluster populations at low redshift
\citep[e.g. de Propris et al. 2004,][]{Margoniner2001,Ellingson2001}, such that
bluer objects reside in the outer parts of clusters. The scatter
of the observed CMR would therefore depend on the maximum
cluster-centric radius sampled in the observations. In addition,
since the observations are not derived from total colours, it is
possible that colour gradients are correlated with deviations from the
true total-colour CMR in such a way that aperture colours appear to have
smaller scatter. For these reasons, we do not find the discrepancy
between the model and observed scatters particularly compelling.

We find that the model predicts a significant correlation between colour and
luminosity. An important
question is \emph{how} the model CMR is generated at present day.
Clearly, in a hierarchical merger scenario age is expected to play
a part in generating any such sequence. It is therefore crucial to
disentangle the effects of age and metallicity and determine how
much of the correlation is generated by a variation
in age and how much by a variation in metallicity with luminosity.

\begin{figure}
\begin{center}
\includegraphics[width=3.4in]{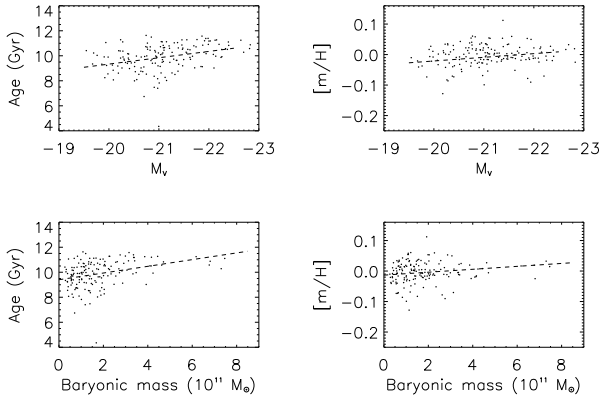}
\caption{\small{Variation of mass-weighted average ages and 
metallicities with absolute V-band luminosity and baryonic mass.}}
\label{fig:t_Z_luminosity}
\end{center}
\end{figure}

\begin{figure}
\begin{center}
\includegraphics[width=3.4in]{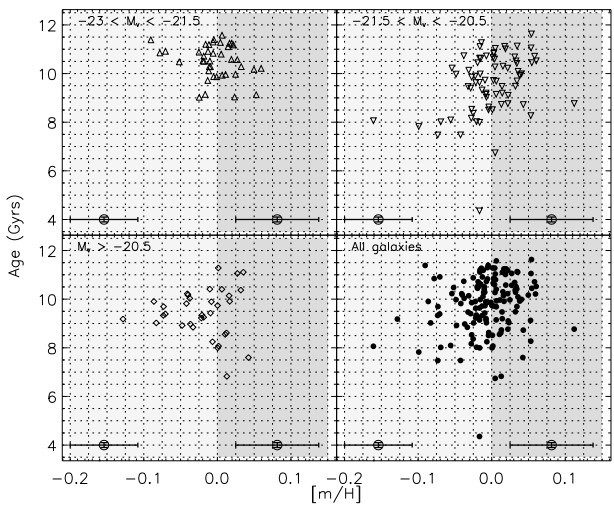}
\caption{\small{Age-metallicity parameter space in model cluster
    ellipticals split into luminosity classes.
    UPPER LEFT: High luminosity model ellipticals ($-23.0 < M_{V} <
    -21.5$); UPPER RIGHT: Intermediate luminosity model ellipticals ($-21.5 < M_{V} <
    -20.5$); LOWER LEFT: Low luminosity model ellipticals ($M_{V} >
    -20.5$); LOWER RIGHT: All model ellipticals. The horizontal error bar shows the \emph{maximum}
    average metallicity error for model galaxies with sub-solar and super-solar metallicities, calculated
    by considering the half-widths of the metallicity bins in the
    model. The vertical error bar shows the maximum error in the ages
    the model ellipticals.}}
\label{fig:t_Z_grid}
\end{center}
\end{figure}

Figure \ref{fig:t_Z_luminosity} shows the variation in the mean 
ages and metallicities of the model cluster
ellipticals with absolute V-band luminosity. We also show the
age-metallicity parameter space for these model galaxies in Figure \ref{fig:t_Z_grid}.
We should note here that the metallicity resolution in the model is
low, with stellar mass resolved only into five metallicity bins in
the range $-1.3<[m/H]<0.5$. We have indicated the \emph{maximum}
average metallicity error for model galaxies with sub-solar and super-solar metallicities in Figure
\ref{fig:t_Z_grid}, by considering the half-widths of our
metallicity bins. The resolution in age, by comparison, is
exteremely good (0.1 Gyr), as indicated by the small age
error bars.

The model predicts a gradient both in the age-luminosity and
metallicity-luminosity relations, so that larger ellipticals are
both older and more metal-rich. We note first that contrary to
previous studies \citep[e.g.][]{KC98},
higher mass (luminosity) ellipticals are predicted to have
\emph{larger} mean ages, in agreement with observational evidence
\cite[see e.g.][]{Trager2000a,Trager2000b,CRC2003}. Current understanding 
of cooling flows in clusters is poor. Models suggest
that if large inflows 
of cold gas are allowed at the centre of virialized DM haloes,
it is impossible to prevent a large fraction of this material from
forming stars \citep[e.g.][]{Cole2000}, which are not observed at the
present epoch. To prevent 
this, one has to \emph{reheat or keep gas hot} in massive 
DM haloes. Various authors have tackled
this problem in different ways. \citet{KWG93}, for instance,
prevented cooling from taking place in DM haloes with circular velocities of
350 km/s and above. GALICS takes advantage of the observed correlation between AGN and
bulge mass \citep{Magorrian98} 
and assumes that AGNs are efficient enough to completely halt cooling 
flows as soon as the bulge which harbours them 
reaches a critical mass of $10^{11}M_\odot$. This coupling 
between AGN feedback and bulge mass prevents star formation early enough 
in large elliptical galaxies to allow them to grow solely through mergers
of gas-poor progenitors. Thus, although
galaxies with a larger mass experience their last merging events
at a later time than their less luminous counterparts, the small gas
fraction at these last-merger epochs prevents any substantial
production of young stars from merger-driven starbursts. Therefore,
although more massive galaxies are
\emph{dynamically younger}
based on their \emph{merger record}, their stellar populations
are, nevertheless, \emph{older}. The average predicted age of a
cluster elliptical is approximately 9.8 Gyrs, and the scatter in
age increases towards the low mass end, in agreement with recent
observational studies in clusters such as Virgo \citep{CRC2003}.
The average metallicity is approximately solar and the gradient in
metallicity is modest, also in general agreement with recent
spectroscopic studies of nearby cluster environments \citep{CRC2003}.

A comparison with simple stellar population (SSP) models \citep{Yi2003} shows
that roughly half of the CMR slope is generated by the
age-luminosity gradient, with the rest attributable to the
metallicity-luminosity gradient in the model sample. Clearly, the
age and metallicity gradients \emph{complement} each other in this
model, in contrast to \citet{KC98} where the
\emph{anti-correlation} between age and luminosity required a
large compensating metallicity gradient (generated through high metal
yields) to produce a CMR that was consistent with the BLE92 observations.

\begin{figure}
\begin{center}
\includegraphics[width=3.4in]{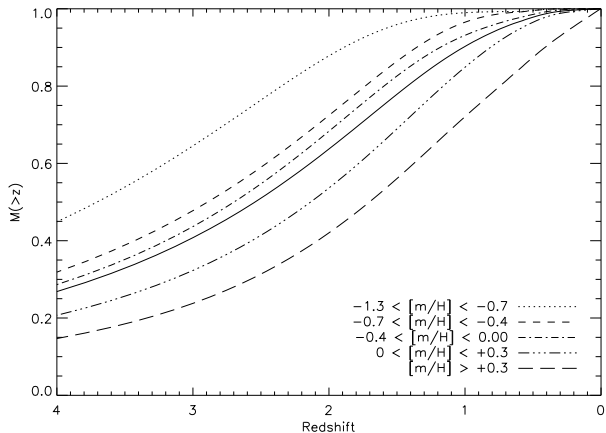}
\caption[Star formation history of cluster
ellipticals]{\small{Stellar mass fraction formed \emph{at or
before} a given redshift. The solid line shows the cumulative 
mass fraction for all stellar mass. The other curves represent stellar
mass in five different metallicity ranges.}}
\label{fig:Bulk_SFH}
\end{center}
\end{figure}

It is clearly beyond the scope of this paper to do a detailed comparison 
of how feedback is treated in GALICS and the specific model of
\cite{KC98}. However, we note that there exists at least two main 
differences:

\begin{enumerate}
\item Cosmological models - \citet{KC98} adopt the SCDM cosmology
  while GALICS adopts the $\Lambda$CDM cosmology.
\item  GALICS derives feedback directly from the mass locked up in the 
spheroidal component of the galaxy, while \citet{KC98} use the 
velocity dispersion in dark matter haloes alone to stop the cooling.
\end{enumerate} 

The first point implies that in GALICS, structures of a given mass will be 
assembled earlier on average than in \citet{KC98}. The second point means
that gas does not cool onto a 
spiral galaxy which sits in a halo with circular velocity greater than
or equal to 350 km/s in the \citet{KC98} model whereas it does in
GALICS, provided the spiral does not possess a massive bulge. Feedback in
GALICS is explicitly linked to the mass build up of 
spheroids, which in turn is correlated to the mass (velocity dispersion) 
build up of the host dark matter halo. However this latter correlation 
need not be linear, since the mass build up of spheroids depends on local 
physics (e.g. disk instabilities and mergers). We attribute the
differences in the results of \citet{KC98} and GALICS to these two
factors (cosmology and feedback modelling) but note that there may be
other factors in the details of the modelling that might cause
discrepancies in the relationship between age, metallicity and
luminosity in these two models.

Figure \ref{fig:Bulk_SFH} presents the bulk cumulative SFH of the
model cluster ellipticals. The SFH is shown both split into the
five GALICS metallicity bins and considering all stellar mass. The
cumulative SFH shows that $\sim$ 10 percent of the total stellar
mass (solid line) was formed after $z=1$, with $\sim$ 65 percent
and $\sim$ 40 percent already in place at $z=2$ and $z=3$
respectively. The bulk
SFH is quasi-monolithic because the low cold gas fraction at low
redshifts ($z<1$) means that merger-driven star formation does not
produce substantial amounts of stellar material. This enables the
model elliptical CMR to maintain its slope and small scatter upto high
redshifts (Section \ref{sec:redshift_evolution}).

\section{Evolution of the CMR with redshift}
\label{sec:redshift_evolution} 
\begin{figure}
\begin{center}
\includegraphics[width=3.4in]{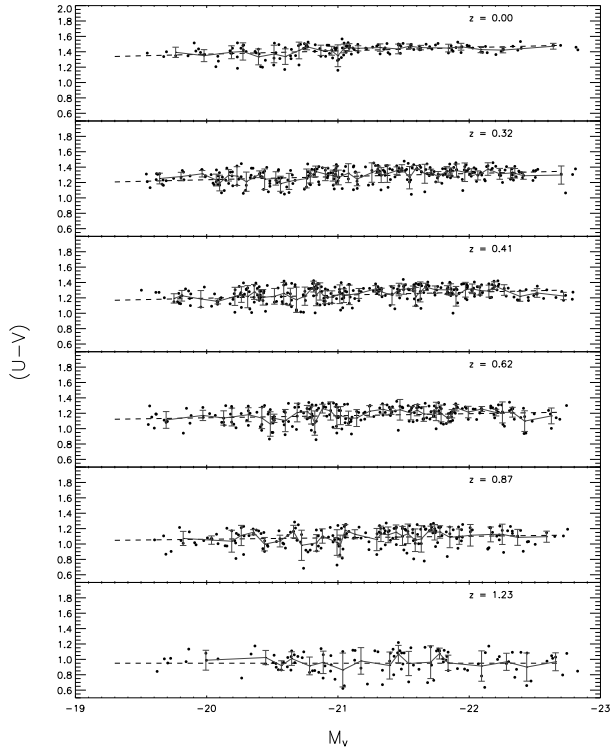}
\caption{\small{Predicted redshift evolution of the model CMR from
    present-day to $z=1.23$, which is roughly the redshift
    limit of current observational evidence on early-type cluster
    galaixes. Also shown is a linear least-squares fit (dotted line)
    and a progressive one-sigma fit, with the error bars indicating the
    local spread of points about the mean relation.}}
\label{fig:CMR_evolution}
\end{center}
\end{figure}

\begin{figure}
\begin{center}
\includegraphics[width=3.4in]{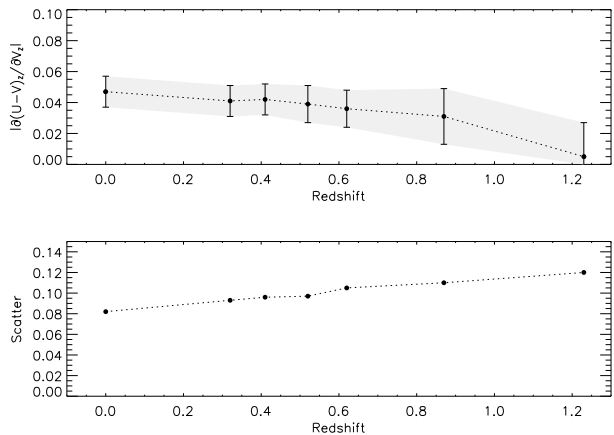}
\caption{\small{Redshift evolution of the slope and 
scatter in the model (U-V) vs. V CMR. Although the evolution in the
slope is zero within the errors in the range $0<z<0.8$, the change
in the slope from the value at present day becomes appreciable at
$z=1.23$.}}
\label{fig:CMR_evolution_values}
\end{center}
\end{figure}

We now check if it is possible to
reconcile the model CMR with observational data at various
redshifts. Figure \ref{fig:CMR_evolution} shows the predicted evolution of
the model CMR from present day to a redshift of 1.27, which
is roughly the redshift limit of current observational evidence on
early-type cluster galaxies \citep{VD2001b}. As
before, the dotted line displays a linear least-squares fit and we
also show a progressive one-sigma fit to the sample, with
the error bars indicating the local spread of points about the
best-fit relation. Figure \ref{fig:CMR_evolution_values} traces 
the evolution of the slope and the
scatter. The shaded region denotes the
area enclosed by the predicted slopes and their associated errors.

We note that the definition of a cluster elliptical will change
with increasing redshift. We assumed in our analysis of present-day 
cluster ellipticals that DM haloes with a mass equal to or
greater than $10^{14} M_{\odot}$ host regions of highest baryonic
density and therefore galaxy clusters. However, since DM haloes
are being steadily formed through time, maintaining a hard mass
cut-off of $10^{14} M_{\odot}$ for all redshifts would not be
correct. To make this definition consistent with changing
redshift we take into account the accretion history of DM haloes
in the model. We first compute an average accretion history of the
present-day DM haloes with masses of $10^{14} M_{\odot}$ and above
as a function of redshift. At each redshift we then define a
\emph{cluster hosting DM halo} as one whose mass is equal to or
exceeds the value given by the average accretion history. We note
that our values are consistent with \citet{Vandenbosch2002} who
provides theoretical prescriptions
for computing universal DM mass accretion histories.

We see from Figure \ref{fig:CMR_evolution_values} 
that there is gradual
evolution in the CMR slope, although in the range $0<z<0.8$ the
evolution in the slope is zero within the errors. However, once
we move out to $z=1.23$ the change in slope is appreciable
compared to the value at present day. Within the errors, we see
that at high redshifts (e.g. $z=1.23$) the CMR loses any
detectable slope, partly because the expected increase in the
scatter masks any correlation that may be present.

\begin{figure}
\begin{center}
\includegraphics[width=3.4in]{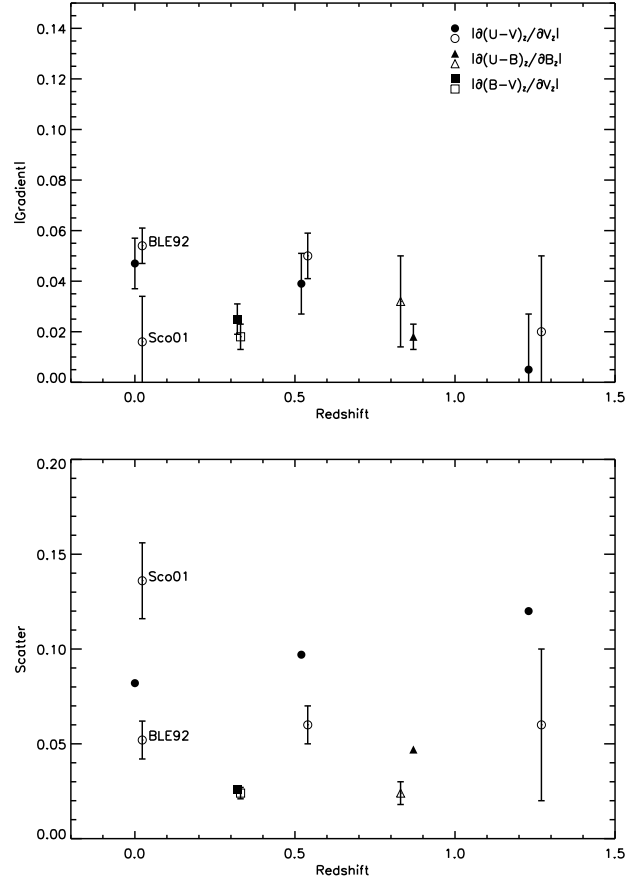}
\caption{\small{CMR evolution with redshift. NOTE: We show the
    properties of CMRs in
    \emph{three different colours} as given in the relevant studies: $(U-V)$
    data are marked as circles, $(U-B)$ data are shown as triangles
    and $(B-V)$ data are shown as squares. Filled symbols represent
    model values and open symbols represent observational results. We
    apply the 30 percent correction for colour gradients derived by
    \citet{Bower98} to studies that have used fixed apertures. The observational data
    from left to right are taken from: \citet{BLE92} (marked),
    \citet{Sco2001} (marked), \citet{VD98}, \citet{Ellis97},
    \citet{VD2000}, \citet{VD2001b}. We do not transform all results
    to a single colour because this requires an assumption of the
    template used to perform the transformation which may introduce
    additional uncertainities into the comparison.}}
\label{fig:cmr_compare}
\end{center}
\end{figure}

In Figure \ref{fig:cmr_compare} we put the evolution of the
predicted CMR in the context of observational evidence. We use a
variety of sources who have explored the CMR in various colours. We
apply the 30 percent correction for colour gradients derived by
\citet{Bower98} to studies that have
used fixed apertures. We find that the slopes of the model and observed CMRs match
well within the errors at all redshifts. In
particular, we note that \citet{VD2001b} reported a slope at
$z=1.27$ that was significantly lower than the BLE92 value at
present day. This suggests that the CMR slope does indeed
decrease, in agreement with the expectations of a hierarchical
merger scenario. The values for the model scatter are also
fairly consistent with the observations, given the previous discussion
in Section 2 regarding possible reasons for the discrepancy between
the model and observed scatter at $z=0$. However, we should note that
the \emph{tightness} of the predicted CMR (especially at the high
luminosity end) seems larger than what appears in observational studies. The model
ellipticals do occupy the \emph{red} part of the sequence (shaded region
in Figure \ref{fig:cmr_compare}), with a scatter to bluer colours which
increases with redshift. However, comparing our results at $z \sim
0.8$ to, for example \citet[][Figure 8]{VD2000}, we find that at comparable
redshift, the observed elliptical CMR is tighter than our model
predictions, although outliers do exist in the observed elliptical
sample.

\section{Progenitor bias}
When comparing the slope and scatter of the CMR at various
redshifts, we should ideally sample the \emph{same stellar
mass at every redshift}. Only then are the slope and scatter
truly meaningful tracers of the star formation history of the
\emph{daughter} mass seen today. However, an unavoidable result of
the merger paradigm is that since early-type systems form through
the amalgamation of late-type units, a progessively \emph{larger}
fraction of the stellar mass we see today in cluster ellipticals
is locked up in late-type (spiral and irregular) units at higher 
redshifts. Hence the early-type systems at high redshift 
form a progressively narrower subset of the progenitors of present-day
elliptical systems.
Consequently, by not taking into account these late-type
progenitors we introduce a bias in the CMR, mainly in terms of the
observed scatter. In this section we quantify the effect of this
\emph{progenitor bias} \cite[see also][]{VD2001b}. Although tracing an astronomical object
back through time is impossible observationally, it becomes a
simple exercise within the model.

\begin{figure}
\begin{center}
\includegraphics[width=3.4in]{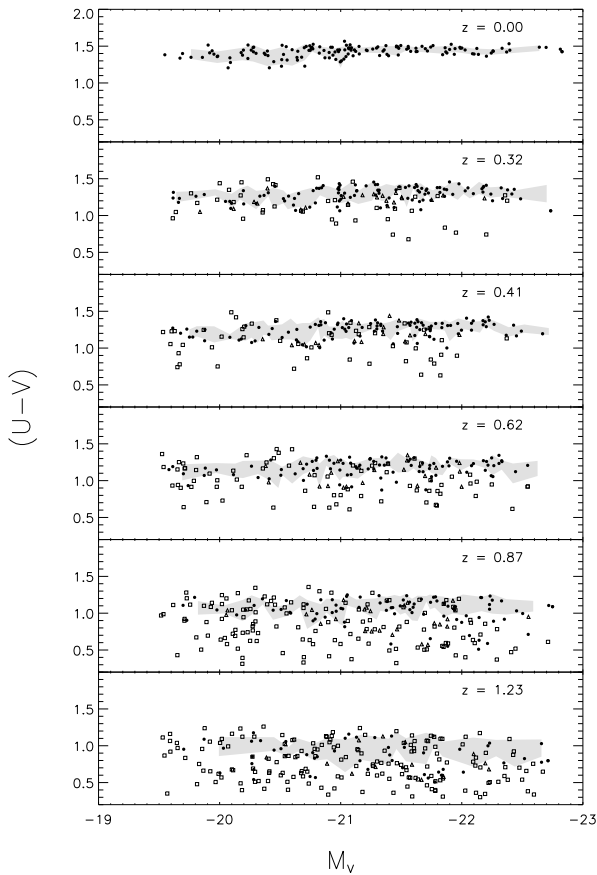}
\caption{\small{Progenitor bias: filled circles
are ellipticals,
triangles are S0s and open squares are late-type systems (spirals and
irregulars). All
galaxies are progenitors of the galaxies at $z=0$. The
shaded region indicates the mean \emph{elliptical-only} relation
and its associated errors taken from Figure
\ref{fig:CMR_evolution}.}}
\label{fig:progenitor_bias}
\end{center}
\end{figure}

\begin{figure}
\begin{center}
\includegraphics[width=3.4in]{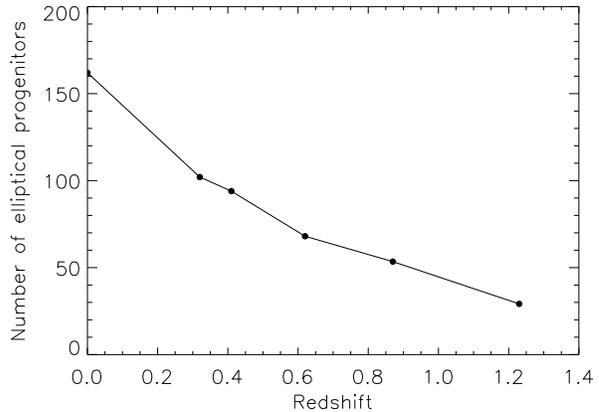}
\caption{\small{The number of fully formed, monolithically evolving elliptical progenitors of
    present-day cluster ellipticals at a given redshift. These
    progenitors do not undergo any further mergers, although quiescent
    star formation continues to $z=0$. The remaining mass in the
    progenitor set is hosted by late-type systems.}}
\label{fig:parent_fraction}
\end{center}
\end{figure}

In Figure \ref{fig:progenitor_bias} we restrict ourselves to the
\emph{progenitor set} of present-day cluster ellipticals. We show only
those galaxies (regardless of morphology) that eventually contribute
to the formation of cluster ellipticals which exist at $z=0$. We are
therefore tracing the \emph{same stellar mass} back through time,
regardless of the type of system that hosts it. We find that including
progenitors with S0 morphology does not change
the slope of the CMR. The overall scatter increases slightly 
at higher redshifts ($z>0.62$), although the elliptical-only scatter
agrees, within  errors, to the E+S0 scatter. S0s, however, tend to
contribute more outliers to the CMR at higher redshifts.
Including the late-type progenitors causes the scatter to increase
approximately three-fold
in the range $z>0.8$, compared to the elliptical-only
scenario. This result agrees with CMR observations at high
redshift. For example, \citet{VD2000} found that the elliptical CMR at
$z=0.83$ has a scatter of $\sim$ 0.024 while the scatter for all morphological types
is $\sim$ 0.081 - an approximate 3.5 fold increase. Blakeslee et
al. (2003) noted that for their observed cluster
at $z=1.24$, deriving the scatter without reference to morphology
increases the CMR scatter three to four-fold. A similar increase can be 
estimated from the study by Van Dokkum et al. (2001) of a cluster at 
$z=1.27$ (see their Figure 3).

Figure \ref{fig:parent_fraction} indicates how much of the progenitor
set of present-day cluster ellipticals is composed of fully-formed
ellipticals at any given redshift. It becomes clear
from Figure \ref{fig:parent_fraction} that if we look solely at
the elliptical progenitors of present-day cluster ellipticals we
sample a progressively thinner fraction of the progenitor set at
higher redshift.
Although restricting ourselves to this subset of progenitors \emph{seems}
to give a CMR which maintains its slope and scatter with
redshift (Figure \ref{fig:CMR_evolution}), an
\emph{elliptical-only} CMR can be used to constrain the SFH
of only the part of the stellar mass in present-day cluster ellipticals
that is contained \emph{solely} in early-type systems at any given redshift.
However, since the subset of elliptical progenitors is
\emph{not representative
  of all the stellar material at present day}, we cannot
  use the evolution of an elliptical-only CMR to
  constrain the SFH of the entire stellar mass of present-day
cluster ellipticals.

Figures \ref{fig:CMR_evolution}, \ref{fig:progenitor_bias} and
\ref{fig:parent_fraction} show that the merger paradigm does indeed
expect to have fully formed elliptical galaxies (and therefore a red
sequence) evolving passively at
redshifts where CMR observations have been conducted. However, the
elliptical-only CMR at high redshift does not correspond to the
elliptical-only CMR at present day and comparisons between the two
give a heavily biased picture of the star formation history of
elliptical galaxies and has serious implications for the ability 
of the CMR to discriminate between the monolithic collapse and
hierachical merger paradigms. In essence, the quantity (in this case
the elliptical-only CMR) that
is being used to discriminate between the two formation models is
no longer model independent and therefore loses its usefulness as a
discriminant.

\section{Summary and conclusions}
We have used a semi-analytical hierarchical galaxy formation model
to investigate the existence and evolution of the CMR of
elliptical galaxies in cluster environments. Our analysis shows
that, by constructing a CMR purely out of early-type systems, the
predicted relation agrees well with local observations (after the
fixed-aperture bias has been corrected) and with observations at all
redshifts in the range $0<z<1.27$. Secondly, we have used our analysis
to quantify the issue of progenitor bias and construct the CMR
that could be expected if we could identify all progenitor systems at high
redshift that would eventually form part of a present-day cluster
elliptical. We have also shown that the scatter in this
\emph{all-progenitor} CMR is consistent with the scatter derived,
without reference to morphology, in cluster CMR studies at high redshift.
Thirdly, we have suggested that the elliptical-only CMR is not a useful
discriminant between the monolithic and merger formation scenarios
since it is significantly biased towards the monolithic picture.
Although the merger paradigm satisfies the elliptical-only CMR in
any case and expects to have a monolithically evolving red sequence at high
redshift, restricting our studies to early-type systems does not
provide meaningful information about the true star formation
history of all the stellar mass that is found today in cluster
ellipticals.

The debate regarding these two competing theories of elliptical
galaxy formation still remains an open one. Although there is clear
evidence of interactions, mergers and recent star formation in
early-type systems, a possible caveat is the inability of the merger paradigm
to satisfy the high [Mg/Fe] ratios observed in luminous
ellipticals \citep[e.g.][]{Trager2000a}. These super-solar
abundance ratios indicate a lack of enrichment from Type Ia
supernovae, thereby constraining the duration of star formation
and gas infall to timescales shorter than about 1 Gyr 
\citep[e.g.][]{Matteucci2001,Ferreras2003}. While the CMR has been
used as an indirect tool for constraining the star formation
history of cluster ellipticals, more direct sources of evidence may
be required. If the stellar mass in cluster ellipticals did
indeed form at $z>>1$ then
we should not find any traces of star formation after this epoch,
which in a $\Lambda$CDM universe corresponds to an age of
approximately 10 Gyrs. The merger models do of course predict star
formation right upto the present day and one could assume that
at least a small fraction of the resultant stellar mass could be
locked up in globular clusters, which are the faintest stellar
aggregations that can be accessed observationally.

\begin{figure}
\begin{center}
\includegraphics[width=3.4in]{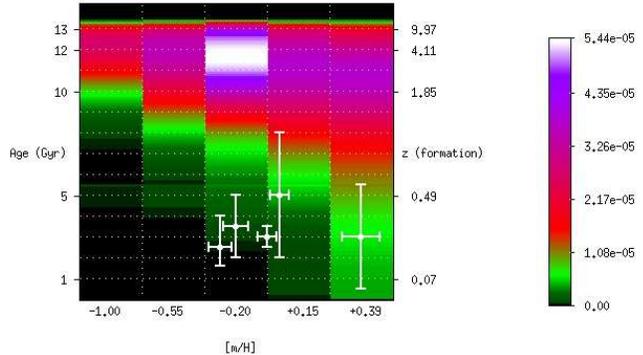}
\caption[Predicted model distribution of stellar mass]
{\small{Predicted distribution of stellar mass contained in cluster
    ellipticals. Overplotted are observations of young globular
    clusters. From left to right -
    \citet{Larsen2003,Strader2003,Goudfrooij2001,KP2002,Yi_2003a}.
    The key indicates the mass fractions corresponding to the colours used in
    the plot.}}
\label{fig:mass_age_distribution}
\end{center}
\end{figure}

Figure \ref{fig:mass_age_distribution} shows the bulk
distribution of stellar mass in present-day cluster ellipticals
predicted by the hierarchical merger paradigm.
One can treat this as a probability distribution of stellar mass,
with the highest intensity areas (see key) indicating ages and
metallicities where most of the stellar mass is likely to be
found. The crucial difference between this model distribution and
a distribution based on the monolithic collapse
model is the presence of \emph{young stars}. Indeed we find that observations of young globular
clusters have been made in elliptical galaxies by a variety of authors
\citep{Goudfrooij2001,KP2002,Larsen2003,Strader2003,Yi_2003a}. 
We indicate these observations in Figure
\ref{fig:mass_age_distribution}. The five data points with error bars
in Figure \ref{fig:mass_age_distribution} show the age/metallicity
properties of young globular cluster populations derived in these studies.
To conclude, we suggest that it seems increasingly likely that the
monolithic collapse picture may
simply be a subset of the merger paradigm and that the dominant
mechanism for the formation of elliptical galaxies is through the
merging of late-type progenitors.

\section*{Acknowledgements}
We are indebted to the referee, Richard Bower, for numerous
suggestions and comments which signifcantly improved the quality of
this paper. We warmly thank Jeremy Blaizot, Roger Davies, Joseph Silk and Sadegh Khochfar for
their careful reading of this
manuscript and many useful discussions. We also thank Seok-Jin Yoon
for constructive remarks related to this work. SK acknowledges PPARC grant
PPA/S/S/2002/03532. This research has been supported by PPARC Theoretical Cosmology Rolling 
Grant PPA/G/O/2001/00016 (S. K. Yi) and has made use of 
Starlink computing facilities at the University of Oxford.

\nocite{Bernardi2003d}
\nocite{dePropris2004}
\nocite{Blakeslee2003}

\bibliographystyle{mn2e}
\bibliography{references}

\end{document}